\documentclass[12pt,preprint]{aastex}
\usepackage{graphics}
\usepackage{epsf}

\begin{document}

\title{Multi-Resolution Radiative Transfer for Line Emission}

\author{M.~Juvela}
\affil{Helsinki University Observatory, T\"ahtitorninm\"aki, P.O.Box 14,
SF-00014 University of Helsinki, Finland}
\email{mika.juvela@helsinki.fi}

\and

\author{P.~Padoan}
\affil{University of California, San Diego}

\begin{abstract}

Radiative transfer calculations are crucial for modelling interstellar 
clouds because they provide the link between physical conditions in a cloud
and the radiation observed from it. Three-dimensional simulations of
magnetohydrodynamic (MHD) turbulence are used to study the structure and
dynamics of interstellar clouds and even to follow the initial stages of core
collapse leading to the formation of new stars. The wide range of size scales
in such models poses a serious challenge to radiative transfer calculations.

In this paper we describe a new computer code that solves the
radiative transfer problem on multi-resolution grids. If the cloud model 
is from an MHD simulation on a regular cartesian grid, criteria based 
for example on local density or velocity gradients are used to refine the 
grid by dividing selected cells into sub-cells. Division can be repeated 
hierarchically. Alternatively, if the cloud model is from an MHD simulations 
with adaptive mesh refinement, the same multi-resolution grid used for the 
MHD simulation is adopted in the radiative transfer calculations.
High discretization is often needed only in a small fraction of the total 
volume. This makes it possible to simulate spectral line maps with good 
accuracy, also minimizing the total number of cells and the computational 
cost (time and memory).

Multi-resolution models are compared with models on regular grids. In the
case of moderate optical depths (e.g. $\tau\sim$ a few) an accuracy of 10\%
can be reached with multi-resolution models where only 10\% of the cells
of the full grid are used. For optically thick species ($\tau\sim 100$),
the same accuracy is achieved using 15\% of the cells. The relation between 
accuracy and number of cells is not found to be significantly different in 
the two MHD models we have studied.
The new code is used to study differences between LTE and non-LTE spectra 
and between isothermal and non-isothermal cloud models. We find significant 
differences in line ratios and individual spectral line profiles of the 
isothermal and LTE models relative to the more realistic non-isothermal
case. The slope of the power spectrum of integrated intensity is instead
very similar in all models.

\end{abstract}


\section{Introduction}

Radiative transfer calculations in three dimensions allow us to study the
effects of density inhomogeneity and complex velocity fields in interstellar
clouds. The first three-dimensional radiative transfer calculations made use of 
non-physical models of the density and velocity fields \citep{park95, park96,
juvela97, pagani98}. A physical description of interstellar cloud structure
and kinematics is provided by three-dimensional simulations of magneto-hydrodynamic 
(MHD) turbulence. Such simulations can presently be run routinely on uniform 
cartesian grids with up to 512$^3$ computational cells (e.g. Li, Norman, Mordecai 
2004) and on multi-resolution grids with adaptive mesh refinement (AMR) methods 
yielding a spatial resolution of many orders of magnitude locally 
\citep{abell02}.

Radiative transfer calculations on a single snapshot of an MHD simulation
are fast compared with the full MHD simulation and are used for 
comparison with observational data. However, each MHD solution can be used
for a number of radiative transfer runs. Radiative transfer can be solved for
different molecular or atomic species and MHD simulations can be scaled to
represent clouds of different size, density, and temperature. In view of the
possibly large number of radiative transfer calculations based on a single 
MHD snapshot, the speed of line transfer calculations is important.

Radiative transfer calculations of molecular lines are often based on Monte
Carlo methods \citep{bernes, choi95, park95, juvela97, hartstein98}. Monte
Carlo methods have the advantage of being simple to implement and easily 
adaptable to different geometries and spatial discretization. Such flexibility
can be important in three-dimensional models where suitable discretization 
(e.g. partial cylindrical symmetry) can significantly reduce run times.  
So far most three-dimensional calculations have been carried out on regular 
cartesian grids. One drawback of Monte Carlo methods is that, due to the 
random sampling, higher accuracy is obtained only with a large increase in 
computational cost. Furthermore, the whole model is usually kept in main 
memory because each of the randomly generated photons can hit any cell in
the model. Finally, it is difficult to carry out Monte Carlo simulations
effectively on parallel computers. 
The lambda operator, $\Lambda$, is at the heart of the radiative transfer 
problem. It maps the source function, $S$, to the average local intensity of
the radiation field, $J$. In Monte Carlo simulation this is accomplished by
simulating model photons and the $\Lambda$ operator is not used explicitly. 

In accelerated lambda iteration (ALI) methods the $\Lambda$ operator is
split into two parts:
\begin{equation}
J = \Lambda S = [ (\Lambda - \Lambda^{*}) + \Lambda^{*} ] S ~ .
\end{equation}
The operator $\Lambda^{*}$ is typically diagonal and represents the
contribution of a cell to the radiation field within the same cell 
\citep[e.g.][]{scharmer81, rybicki91, rybicki92}. Iterations are needed to 
solve for large scale interactions. The convergence of such iterations becomes
faster for increasingly higher optical depths and weaker coupling between
cells. ALI methods have been used also with two-dimensional 
\citep[e.g.][]{auer94, dullemond00} and three-dimensional grids. 

In contrast with Monte Carlo methods, in ALI implementations the sampling 
of the radiation field is usually not random and is based on a fixed set of
characteristics. A characteristic is a line along which the radiative transfer
equation is solved and where the intensity of the incoming radiation is known.  
For long characteristics, integration is done for each cell separately starting
from the cloud surface. However, most three-dimensional implementations are 
based on short characteristics \citep{kunasz88,auer_paletou94,auer94}. 
Short characteristics extend only from one cell to the next one and, on 
cartesian grids, intensity for any given direction may be propagated only one 
cell layer at a time. This helps reduce memory requirements by dividing the 
model cloud into separate layers that are treated sequentially. The 
computational overhead resulting from such division is much smaller than 
the equivalent overhead in Monte Carlo methods. 

The only real difference between Monte Carlo and ALI methods is in the
sampling of the radiation field. 
The so-called accelerated Monte Carlo methods correspond directly to ALI
methods \citep{alma99, hogerheijde00, zaddelhoff02}.  The basic idea in the
core saturation method of \citet{hartstein98} is similar, but uses a
distinction between optically thick line centre and optically thin wings.

Computer resources limit the detail of current models. One way to speed up 
the solution of the radiative transfer problem is to resort to approximations.
For example, \citet{ossenkopf02} combined a local large velocity gradient
(LVG) approximation with an approximation of an isotropic large scale field.
Tests with isothermal, three dimensional MHD models showed the accuracy of the
method to be $\sim$20\% or better while speed-ups were very significant.

Another possibility is to solve the radiative transfer problem without
approximations and to decrease the number of cells in the model. Interstellar
clouds are very inhomogeneous and volume filling factors derived from
molecular line or infrared emission are low, varying from a few tenths in star
forming cores \citep{snell84,greaves92} to less than one percent in molecular 
cloud complexes \citep{falgarone91}. This suggests the emission of many 
molecular species can be predicted with high precision even if high spatial
resolution is only used for some small fraction of the total volume. 
Multi-resolution grids are already used in cosmology \citep{abell02},
in studies of supernovae driven galactic disk fountains \citep{deAvillez01},
to simulate star-forming clouds \citep{klein03} and in studies of 
incompressible MHD flows \citep{Grauer+2000}
thanks to adaptive mesh refinement (AMR) methods. Radiative transfer 
calculations based on solutions of AMR simulations of MHD turbulence in 
interstellar clouds must be performed in a way that takes advantage of
multi-resolution grids.

In this paper we present a computer code that can be used to solve the
radiative transfer problem for line radiation using hierarchically refined
spatial discretization. The implementation is presented in
Sect.~\ref{sect:program}. We compute $^{13}$CO and $^{12}$CO spectra for 
isothermal models, based on the solution of MHD simulations performed
on a regularly spaced cartesian grid. In Sect.~\ref{sect:tests} we study 
the accuracy and the reduction in the required computational resources of 
this multi-resolution code. In Sect.~\ref{sect:comparison} we use the 
new code to study differences between
LTE and non-LTE spectra and between isothermal and non-isothermal cloud
models.  In Sect.~\ref{sect:discussion} e.g. further improvements to the current
code are discussed.

\section{The code} \label{sect:program}

\begin{figure}
\epsscale{0.60}
\plotone{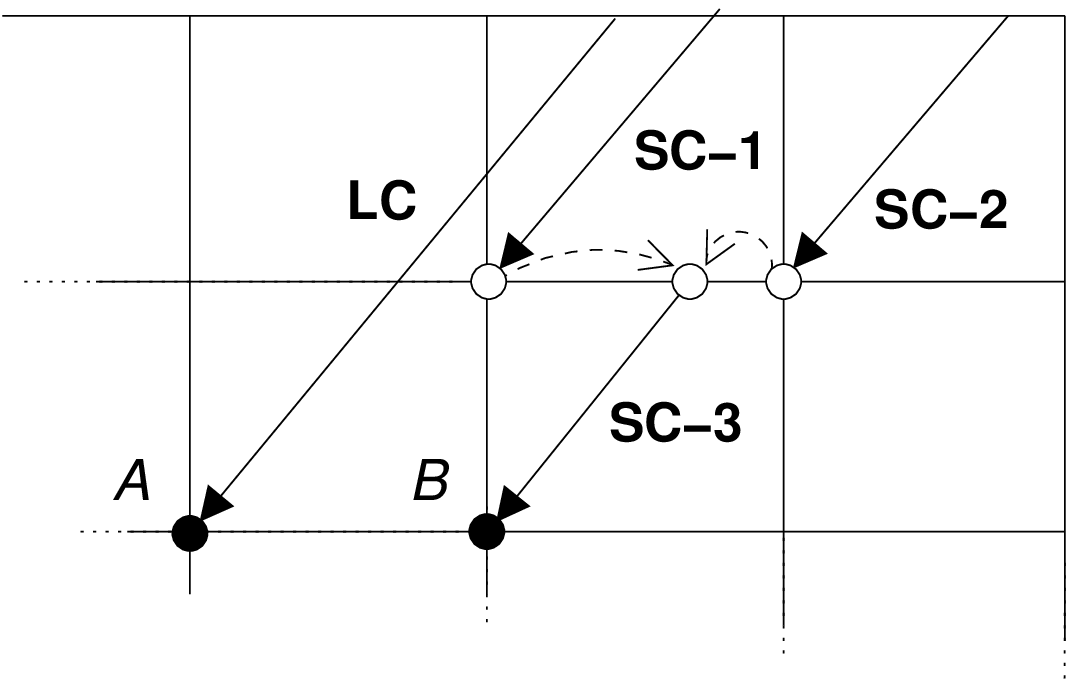}
\caption[]{
Two schemes for computing intensity at a grid position. With a long characteristic
(LC) calculation is done by integrating directly from the cloud surface to a grid
position $A$. Alternatively, the intensity can be propagated one grid layer at a
time using short characteristics (SC).  In the latter case interpolation
(dashed arrows) is used to determine the intensity at the starting position of the
third SC that ends at the grid position $B$.
} \label{fig:lc_sc}
\end{figure}

\subsection{Background: long and short characteristics}

The lambda operator $\Lambda$ can be implemented numerically in different ways,
depending on how the intensity averaged over the whole solid angle is computed
at each position in the cloud. The simplest way is to use long characteristics (LC)
that extend directly from the cloud surface to a grid point inside the cloud (see
Fig.~\ref{fig:lc_sc}). The radiative transfer equation is integrated along a LC
giving the intensity in one direction. The angle-averaged intensity is then 
obtained by repeating the calculations for a sufficiently large number of LCs 
pointing in different directions. These calculations are repeated independently 
for each grid position.

Calculations with LCs contain redundancy. A LC is used only to estimate
intensity at its end point and not for any of the other cells it may
cross. In short-characteristic (SC) methods this redundancy is removed. A short
characteristic extends only between two layers of grid points. The grid must
be swept in an ordered fashion so that intensities are always known for the
previous layer. When a short characteristic is formed and
extended backward it usually ends up between grid points in the previous
layer. Therefore, intensity at the starting position is first interpolated and
integration along the SC then propagates the intensity information
to the next layer (see Fig.~\ref{fig:lc_sc}).

In the case of SCs, consecutive interpolations lead to diffusion, that is 
the intensity propagates not only in the direction of the SCs, but also 
perpendicular to that. The accuracy of the interpolation is essential to 
minimize this effect. 
\citet{auer_paletou94} note that one should use at least second
order schemes and additionally guard against negative values introduced by
interpolation.

\subsection{The basic scheme}

\begin{figure}
\epsscale{0.60}
\plotone{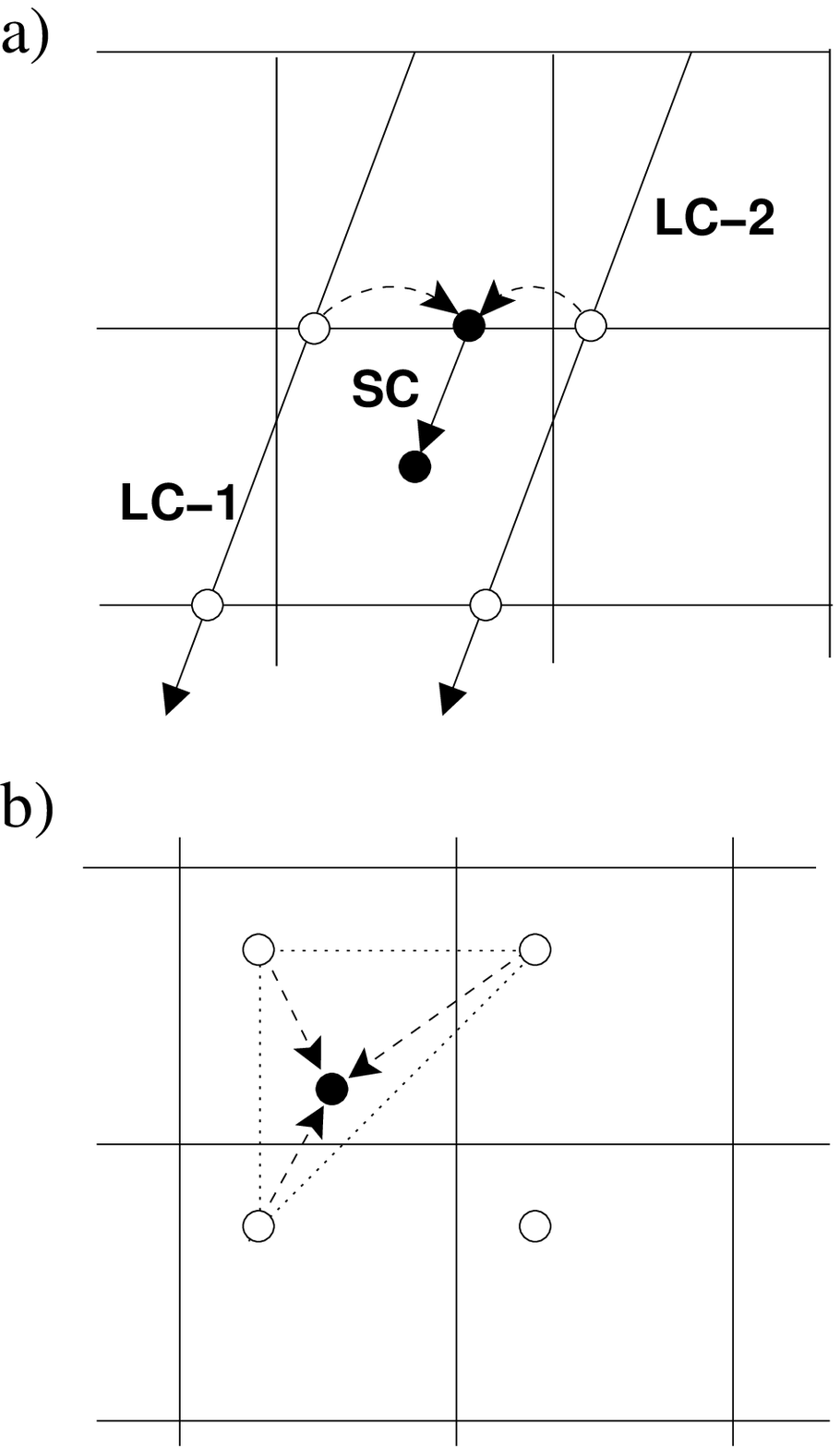}
\caption[]{
Our scheme for the propagation of intensity information. The grid
denotes {\em cells} at the centre of which intensity is computed. Long
characteristics (LC) are used to integrate the intensity, one cell layer 
at a time, so that accurate intensities are known at the cell faces (open circles).
Interpolation (dashed arrows) is used to derive the intensity at the starting
point of a short characteristic (SC) used to integrate the intensity to the 
centre of a cell (filled circle). 
The lower frame shows the situation from above: interpolation is done
using the closest three LCs. 
} \label{fig:method}
\end{figure}

In our scheme each grid position represents the centre of a cubic cell and we
combine the use of long and short characteristics (see Fig.~\ref{fig:method}).
At any given time only intensities for one cell layer are kept in memory.
LCs are initially created at intervals corresponding to the cell size. The 
intensity is propagated along these LCs. The computation proceeds one layer at 
a time in order to avoid any redundancy. In order to determine the intensity at the 
cell centre, a short characteristic is formed. At the starting point of a SC 
the intensity is linearly interpolated using the values at the positions of 
the three closest LCs.

Intensities along LCs are the result of direct integration along these
lines. This prevents the appearance of any cumulative errors since intensity
values do not depend on any interpolation. Interpolation is performed only at
the starting position of each SC. The three closest LCs surrounding the SC are
selected (see Fig.~\ref{fig:method}b) and intensity at the start of the SC is
obtained with linear interpolation. This operation is much less critical than
in pure SC schemes where interpolation would affect all intensities in the
downstream direction. In our case interpolation errors are not propagated
beyond the current cell. One could even take the intensity of the closest LC
without any interpolation. Resulting errors would still correspond only to
differences in the radiation field at scales below the size of an individual
cell. In integration along SC, the intensity is split into an external part
and a part caused by the cell in question, thus leading to an ALI method with
a diagonal operator $\Lambda^{*}$.

Previous SC codes were designed to give good accuracy in the case of
smoothly varying source functions. For fractal interstellar clouds the
underlying density distribution is not smooth and there are no guarantees of
accuracy of high order interpolation. We use only linear interpolation
when intensities at the starting positions of the SCs are derived. This is
possible since we use LCs to carry intensity information through the whole
cloud. We take a simple approach also regarding the integration along
the characteristics. As in the Monte Carlo code, we assume the source function 
and the optical depth are constant within individual cells. 

Our code handles three dimensional clouds where basic discretization is
done according to a cartesian grid. The program is optimized for low memory
requirement so that at any given time only data for one layer of cells are
kept in memory. These data include intensities at the positions where LCs
enter the current layer of cells. The intensity is interpolated to the starting 
position of a SC and the integration along the SC gives the intensity at the cell 
centre. These intensities can be saved to an external file or, if the model size
allows it, they can be directly added to counters that are later used to calculate
intensities averaged over solid angle. Once LCs have been integrated to the
boundary of the next cell layer, data for the previous layer can be removed and
data for the next layer are read from external files. This is repeated
until all cell layers have been processed. Calculations are repeated for a
number of directions. Finally, angle-averaged intensities are computed for each 
cell. These steps are carried out for each transition separately. It would be more
efficient to calculate all transitions simultaneously, but
that would also increase the memory requirement by a factor equal to the number
of transitions simulated. Once intensities in the cells are known, new
estimates of level populations can be computed from the equilibrium equations. 
Updated source functions and optical depths are then computed and the whole
procedure is repeated until changes in the level populations between consecutive
iterations are below the required accuracy.

\subsection{Hierarchical refinement}

A multi-resolution model cloud may be directly provided as the solution
of an AMR simulation of MHD turbulence. If the model cloud is instead
the solution of the MHD equations on a regular cartesian grid, a 
multi-resolution grid can be obtained prior to the radiative transfer
calculations. We refer to the lowest resolution root grid as the level
1 grid. At the level 1, each cell can be divided into eight level 2 cells 
and this division can be repeated for any of the sub-cells. High levels
of discretization are used in regions where high accuracy is required
to compute the spectral line maps, such as in regions of high density
or large velocity gradients. The specific refinement criterion can be fit
to the specific model and needs.

The method described in the previous chapter is used to calculate the
intensities in the multi-resolution models. Because the code is designed to
minimize the memory requirement, only one layer of level 1 cells and their
possible sub-cells are kept in memory. 
For each cell touching the upper boundary of the layer there should be
exactly one LC. Cells at level 1 are handled row by row starting with the
upstream direction as defined by the orientation of the LCs. Within each level 1
cell all sub-cells and the corresponding LCs are processed recursively in the
same order. Each LC is integrated down to the next boundary between level 1
cells. After these calculations intensities are known in each position where a
LC hits a new cell and later these values are interpolated to give
intensities at the start positions of the SCs.

\begin{figure}
\epsscale{0.70}
\plotone{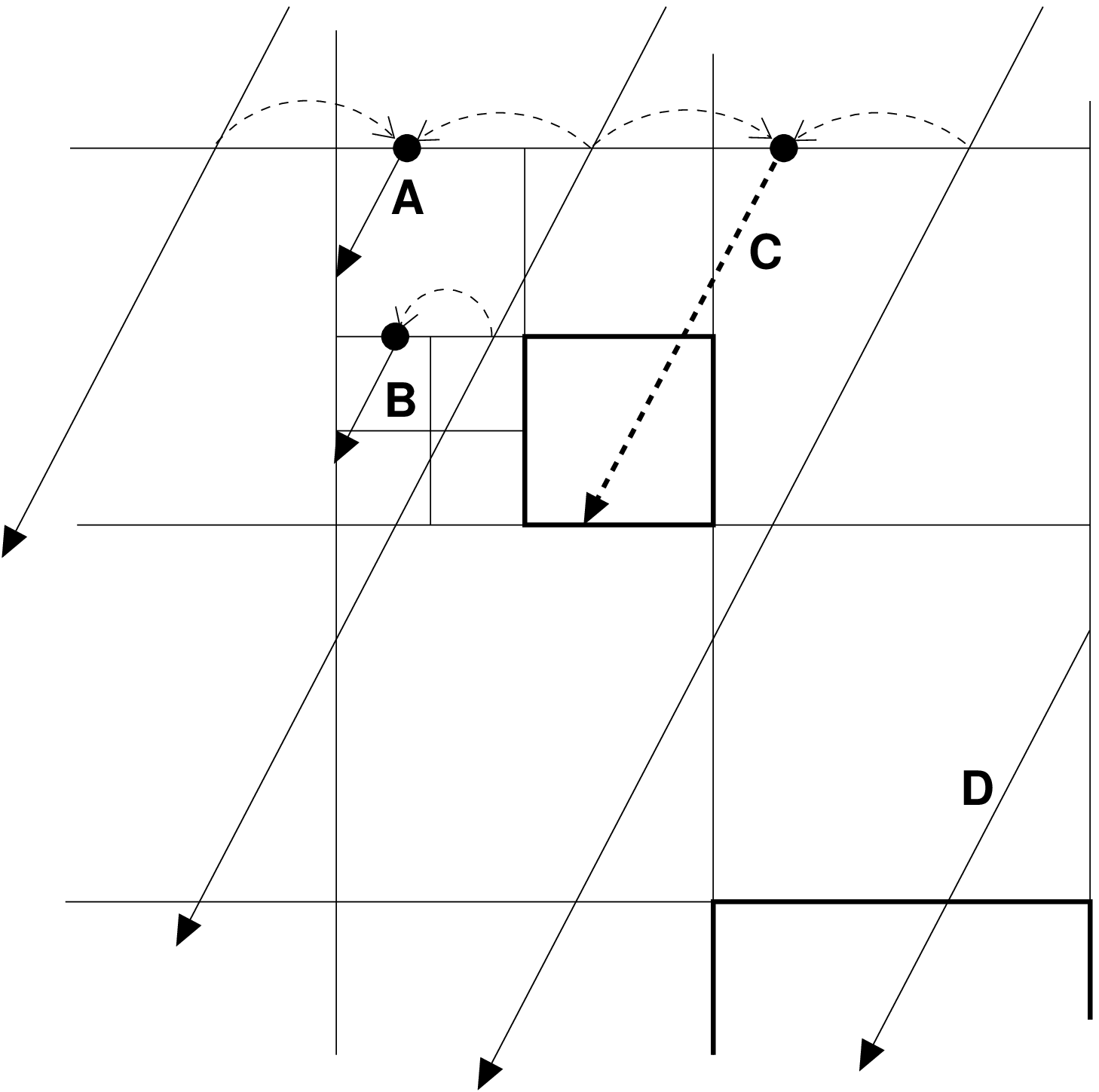}
\caption[]{

Examples (in 2D) of the creation and deletion of LCs as the cell
discretization changes along their path. Solid arrows stand for LCs and
dashed arrows denote interpolation. The four long arrows are level 1 LCs
that start at the cloud surface and are never deleted. $A$ is a new level 2 LC
that is created by interpolation of level 1 LCs and is deleted when it enters a 
level 1 cell. The intensity at the starting point of the level 3 LC $B$ is taken 
directly from the nearest LC because level 2 LCs are not present on both sides. 
The LC $C$ is created only later, at the time of the creation of SCs, when it is 
noticed that the top of the cell (thick box) is not hit by any of the existing LCs. 
The LC $C$ is traced back to the previous cell boundary where interpolation takes 
place. Finally, the LC $D$ is also created later and it is traced back to the 
cloud surface.
} \label{fig:lc_lc}
\end{figure}

In multi-resolution models new LCs are created or destroyed as discretization
changes (Fig.~\ref{fig:lc_lc}).  If a LC enters a smaller cell, new LCs must be
created, one for each of the neighbouring sub-cells. Three new LCs must be
created if change is from level 1 to level 2 (or from level 2 to 3); 15 if
change is directly from level 1 to level 3. Intensities for the new LCs are
interpolated between the nearest existing LCs. A more simple procedure would be 
to copy the intensities from the nearest LC and errors would again correspond only to
intensity variations at distances smaller than the size of the cell from
which the LC come. Currently this procedure is used only when interpolation is not
possible, for positions without suitable LCs on opposite sides (see
Fig.~\ref{fig:lc_lc}).  

When a LC comes from a smaller cell into a larger cell it may have to be
removed. For this purpose, a level number is attached to each LC. For example,
if a LC is created entering a level $l$ cell, it will be
destroyed only when it enters a cell with level smaller than $l$. When some of
level $l$ LCs are destroyed one could calculate the average of the remaining LCs
at level $l-1$. Currently unnecessary LCs are only deleted and
the remaining LC at level $l-1$ represents the intensity integrated strictly along
that one line-of-sight. 

The implementation of the program is discussed further in
Appendix~\ref{sect:implementation}.

\section{Tests of the method} \label{sect:tests}

\subsection{Comparison with the Monte Carlo code}

The new code was first tested against our Monte Carlo code 
\citep{juvela97} that has already been tested and has recently been compared with
other radiative transfer codes 
\citep{zaddelhoff02}. Tests of our new code were done using single-resolution
grids and MHD cloud models similar to those used in
\citet{juvela03}. Models represent an interstellar cloud with a diameter of
$\sim$10\,pc, average density 500\,cm$^{-3}$ and temperature of 10\,K.  Two
supersonic models with Mach numbers $M=2.5$ (model $A$) and $M=10$ (model $B$)
were used \citep[for more details see][]{padoan03}.
The resolution is $124^3$ cells. Tests were repeated for $^{13}$CO
and $^{12}$CO so that average optical depths of the lowest transitions ranged
from a few to over one hundred.

For the new ALI code we first used a discretization where all the cells were
subdivided twice and the resulting cells exactly corresponded to the 124$^3$
cell grid of Monte Carlo calculations.  Line intensities from the Monte Carlo
code and the new code agree very well. The difference in average intensity is
less than 1\%, and rms differences computed over maps of 124$\times$124
spectral lines are below $\sim$2\%. The remaining variations can be explained
with sampling errors caused by the finite number of simulated photon packages
in Monte Carlo (800,000 per iteration) and the finite number of directions
($\sim 40$) in the ALI runs. In both codes the cloud description is
practically identical and the cloud consists of cells which are internally
homogeneous. For example, in ALI calculations we do not interpolate source
functions and absorption coefficients when the radiative transfer equation is
integrated along a SC. The main difference between the two methods is the
following. In the Monte Carlo code the local intensity is described through 
the number of absorbed photons, which is the result of integration along random
lines of sight through a cell. In the new ALI code the intensity is calculated
for the central position of each cell. The two results may differ if there is
a large difference between the radiation field at the cell centre and at the
cell boundaries. This is possible only if individual cells are optically
very thick. The small difference in the results from the two codes shows that
this effect is not important for the current models. The small differences
between the maps produced by the two codes are also of a random nature, and do
not correlate for example with the position of dense filaments.

\begin{figure}
\epsscale{0.60}
\plotone{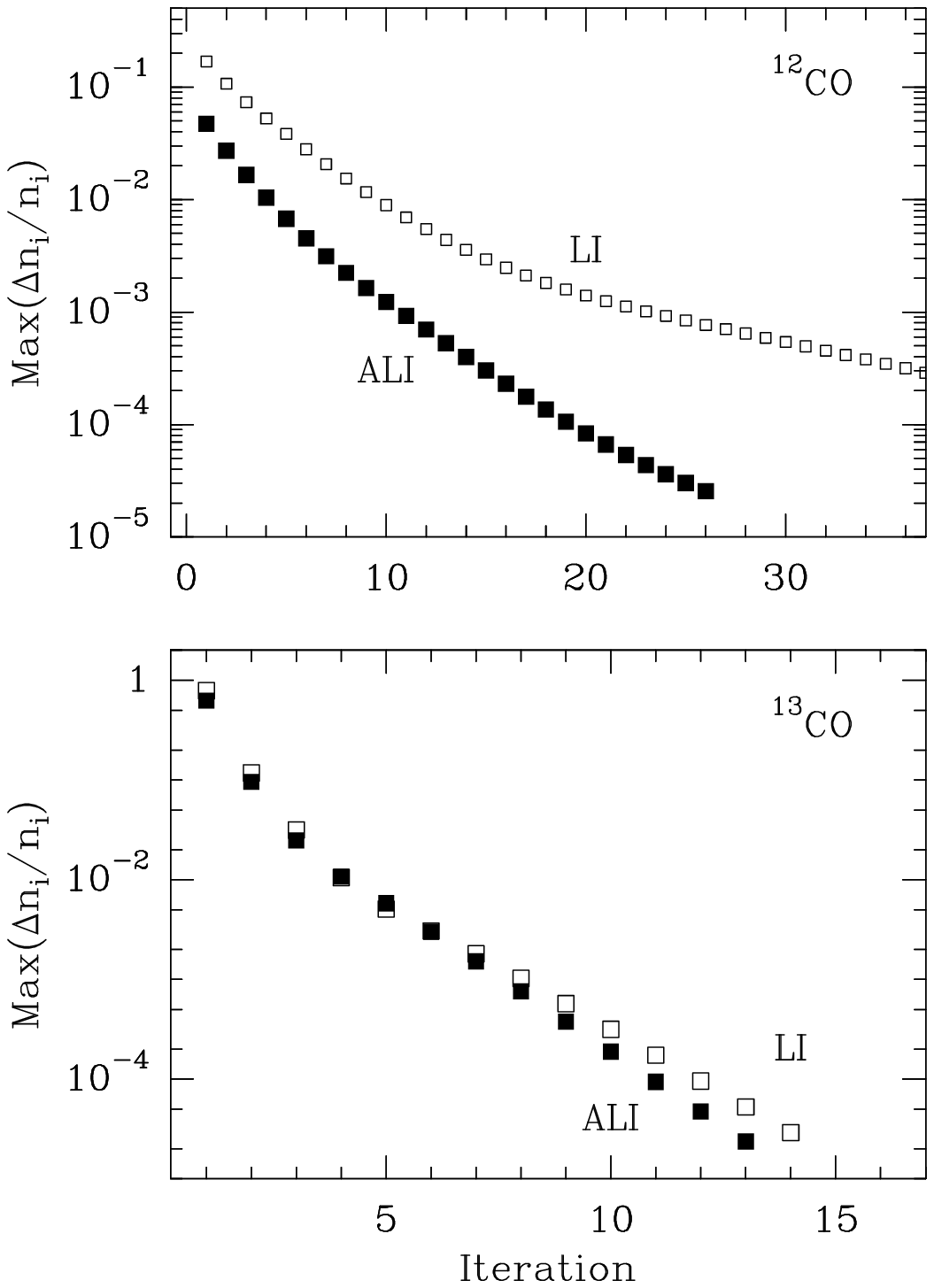}
\caption[]{
Convergence of $^{12}$CO level populations with lambda iteration and
accelerated lambda iteration in the case of the MHD model $A$ with sonic Mach
number 2.5. The maximum relative change in level population, $\Delta n_i/n_i$,
was calculated for transitions up to $J=4$ for each cell with $n>50\,$cm$^{-3}$.
The average of these values is plotted here. The average optical depth for the
transition $J$=1-0 is $\sim 4$ and $\sim$ 124 for $^{13}$CO and $^{12}$CO
respectively.
} \label{fig:convergence_iter}
\end{figure}

Convergence was checked with and without ALI acceleration. All calculations
were started with LTE level populations assuming excitation temperature equal
to kinetic temperature.  This is a better approximation for dense gas than for 
lower density regions where the converge is fast anyway.
Fig.~\ref{fig:convergence_iter} shows the convergence in model $A$.  For
$^{13}$CO the difference between the two methods is small but for 
optically thick $^{12}$CO the ALI method is significantly faster.  Another
noticeable feature is the non-linearity (on a logarithmic scale) of the
convergence. Initial fast convergence is due to the low-density gas
where the relative change in the level population, $\Delta n_i/n_i$, becomes 
small very quickly. At later iterations, the average convergence rate depends 
on gradually more opaque regions where convergence is slower. The ALI method 
improves the convergence in opaque regions and the change
in the overall convergence rate versus the number of iterations is much smaller 
than for the lambda iteration method. However, an inhomogeneous cloud always has 
regions with different convergence rates and the initial behaviour is generally
not a good indicator of the convergence rates after many iterations.  

\subsection{Tests of multi-resolution calculations}

We now report on the comparison between single-resolution (SR) and multi-resolution
(MR) models with three levels hierarchy. In MR calculations part
of the level 1 cells are divided into eight level 2 cells and each of these
may in turn be divided into eight level 3 cells. On lower levels (i.e. for
larger cells) density and velocity are averaged over sub-cells, and the
intrinsic line-width is calculated taking into account velocity dispersion
between sub-cells. Level 3 cells correspond exactly to cells in the SR
models. The comparison between SR and MR models allows us to test the accuracy
of the MR code as a function of the number of computational cells. MR models are
useful only if a reasonable accuracy is achieved with a number of computational
cells significantly smaller than in the SR models. The ratio between the number of 
cells in the MR and SR models, $R$, is roughly proportional to the ratios in 
computational times and in the memory requirements of the two models. It should be 
kept as low as possible, provided the spectral line profiles are computed with 
the required accuracy.

\begin{figure}
\plotone{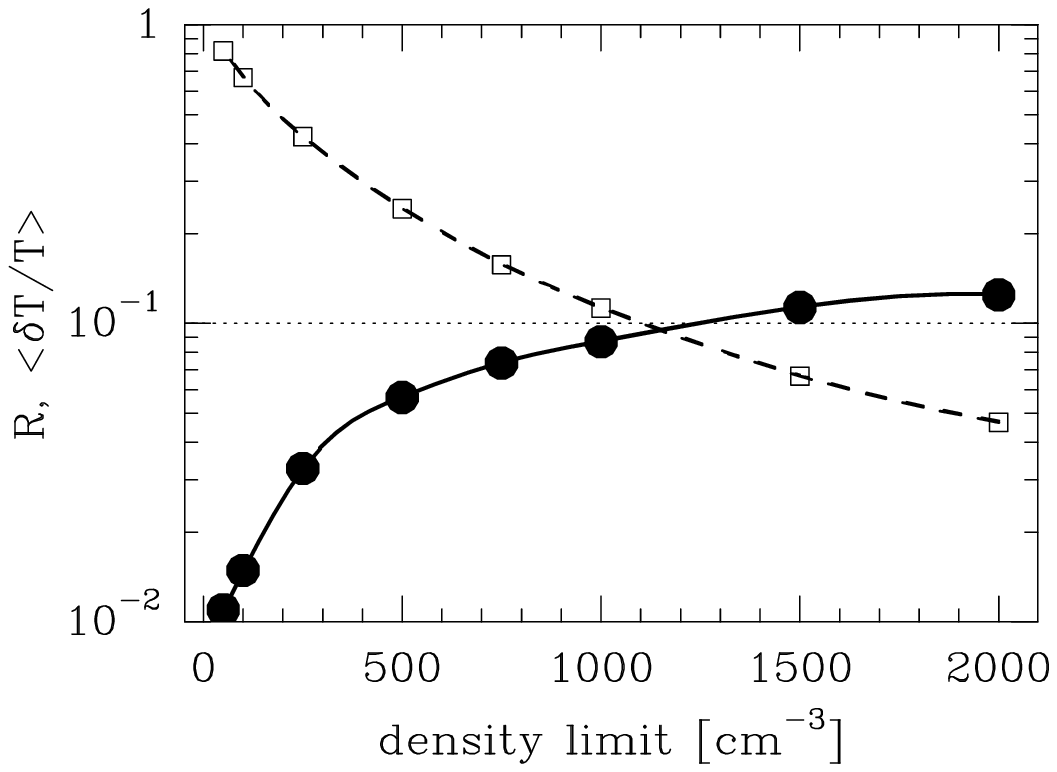}
\caption[]{
Accuracy of multiresolution calculations as a function of the density
threshold for grid refinement. The rms-error in the computed intensities of
the $^{13}$CO(1--0) line (filled circles) and the value of the ratio of the
number of cells in the MR and SR models, $R$, (open squares) are plotted
versus the density threshold for model $B$. The plot shows results for
model $B$ but very similar values are obtained for model $A$.
} \label{fig:2f_accuracy}
\end{figure}

\begin{figure}
\epsscale{0.70}
\plotone{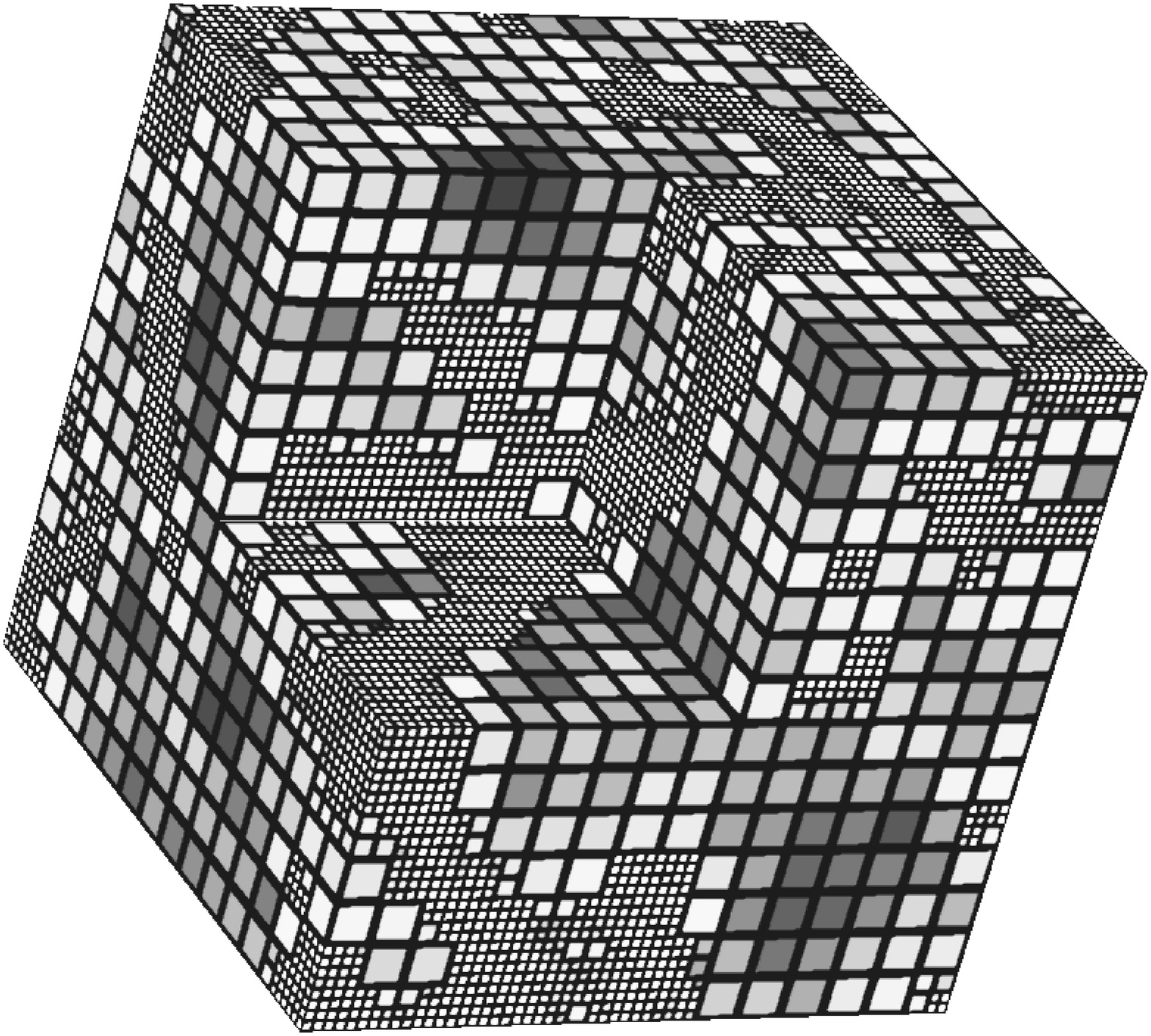}
\caption[]{
Discretization for model $A$ with grid refinement above a density of 500\,cm$^{-3}$. 
For clarity, this figure shows a model with a decreased resolution of 60$^3$ cells. 
One corner of the cloud has been extracted to show the discretization inside the cloud.
} \label{fig:division}
\end{figure}

For the purpose of this comparison we have used a refinement criterion based
on the local gas density. A large fraction of the cloud volume is filled with 
low-density gas. Unless optical depths are very high, only dense parts are relevant 
for a comparison with observational data, while the rest of the volume can be modelled
at a low resolution. In the two MHD models we have used in this work the
average density is 500\,cm$^{-3}$, while peak values are higher by a factor of 33
for model $A$ and a factor of 170 for model $B$. The rms error of the $^{13}$CO line
intensities was computed as a function of the threshold density for refinement.
In model $B$, an accuracy of 10\% is reached with $R \sim 0.1$, that is with 
computations that are one order of magnitude faster than for the SR model (see
Fig.~\ref{fig:2f_accuracy} and Fig.~\ref{fig:division}). In model $A$ the density
contrast is lower but accuracy turns out to be even slightly better.

The accuracy is expected to be better for species that are optically
thinner or have higher critical densities than $^{13}$CO since for these the
emission will be even more concentrated in the densest regions. 
Conversely, for optically thick molecules, one might think that good
accuracy would require higher $R$ values. A larger fraction of cloud volume
becomes relevant for observed lines but, on the other hand, excitation
temperature variations are decreased by line thermalization.  For $^{12}$CO
the average optical depth of the $J=1-0$ line is in our model clouds $\sim
100$.
An accuracy of 10\% is reached for both models with $R\la0.15$. This is
not very different from the value of $R$ required for the $^{13}$CO line.
However, an accuracy of 5\% is achieved for the $^{12}$CO line only with
values of $R$ above 0.5, making the advantage of the MR calculations very
small if high precision is required for such optically thick lines.
By modifying the molecular abundance we calculated a few
models with average optical depths between the previous $^{13}$CO
and $^{12}$CO models. These confirmed a gradual decrease
of accuracy with increasing optical depth.
Higher discretization in optically very thin regions does little to
improve the accuracy of computed spectra. The same seems to apply to very high
optical depths. If one assumes a constant kinetic temperature also the
excitation is in these regions nearly constant and because of the foreground
absorption these regions contribute directly only little to the observed
intensity. 

Apart from the velocity field the main factors affecting the accuracy of
the MR calculations are line optical depth and the inhomogeneity of the cloud
model. Previous tests showed that calculations are feasible for a wide range
of optical depths, at least in the range $\tau \sim$1-100. 
For higher depths one could improve the efficiency by using smaller cell size
on that side of the cloud that is facing the observer - provided that it is
sufficient to calculate the spectra just for one viewing direction. 

The small difference between the two model clouds suggests that accuracy is
not strongly dependent on the density distribution.  However, both models are
rather clumpy and one might ask whether the situation might change in more
homogeneous clouds where observed intensities are again affected by a larger
fraction of the cloud volume. We modified the density contrast of model $B$
and checked the effect this has on the accuracy of $^{13}$CO spectra. The
average density, $n$=500\,cm$^{-3}$, was kept constant while density
variations were decreased by a factor of 2, 5, 10, or 50. Cells above a
density limit of $n$=500\,cm$^{-3}$ were split so that the discretization was
in all cases exactly the same. In the original model the rms error of line
intensities was slightly below 6\% (see Fig.~\ref{fig:2f_accuracy}). The error
increased with decreasing density variations but was still less than 12\% when
density contrast was decreased by a factor of 50.
The result is not altogether surprising. Fine discretization is needed in
order to resolve variations in density and excitation. In a homogeneous cloud
these variations become small and errors resulting from a poor discretization
do not grow very large.

It may be possible to define a better criterion for cell refinement if
information on excitation temperature ($T_{\rm ex}$) and velocity fields is
included. The excitation temperature is the most important variable affecting the
emerging intensity, although its variations mostly follow density
variations. 
Good sampling of the velocity field would be particularly important if we
were interested in line profiles rather than total line intensities.
Velocity field determines the radiative coupling between different
areas and the spatial resolution in regions of high velocity gradients could
have an impact even on line intensities.

As an alternative refinement criterion we set threshold values for
velocity difference, $\Delta V$, and excitation temperature difference,
$\Delta T$, and refine all cells where such differences relative to
neighbouring cells are above a given threshold. 
All combinations of $\Delta V$ and $\Delta T$
thresholds resulted in an accuracy that was always worse or only roughly equal
than with the simple density criterion with equal values of $R$. The $\Delta
V$ and $\Delta T$ limits force the refinement in all regions with velocity and
excitation gradients, even in low density regions which do not contribute much
to the observed spectra. This may explain the relatively poor result. On the
other hand, these tests show that density does indeed provide a good criterion
for hierarchical refinement. With a suitable combination of density,
excitation temperature and velocity criteria one does obtain better results.
The improvement is, however, only $\sim$1\% in accuracy for any given value of
$R$ and in the following we rely only on a density thresholds.

\begin{figure}
\epsscale{0.58}
\plotone{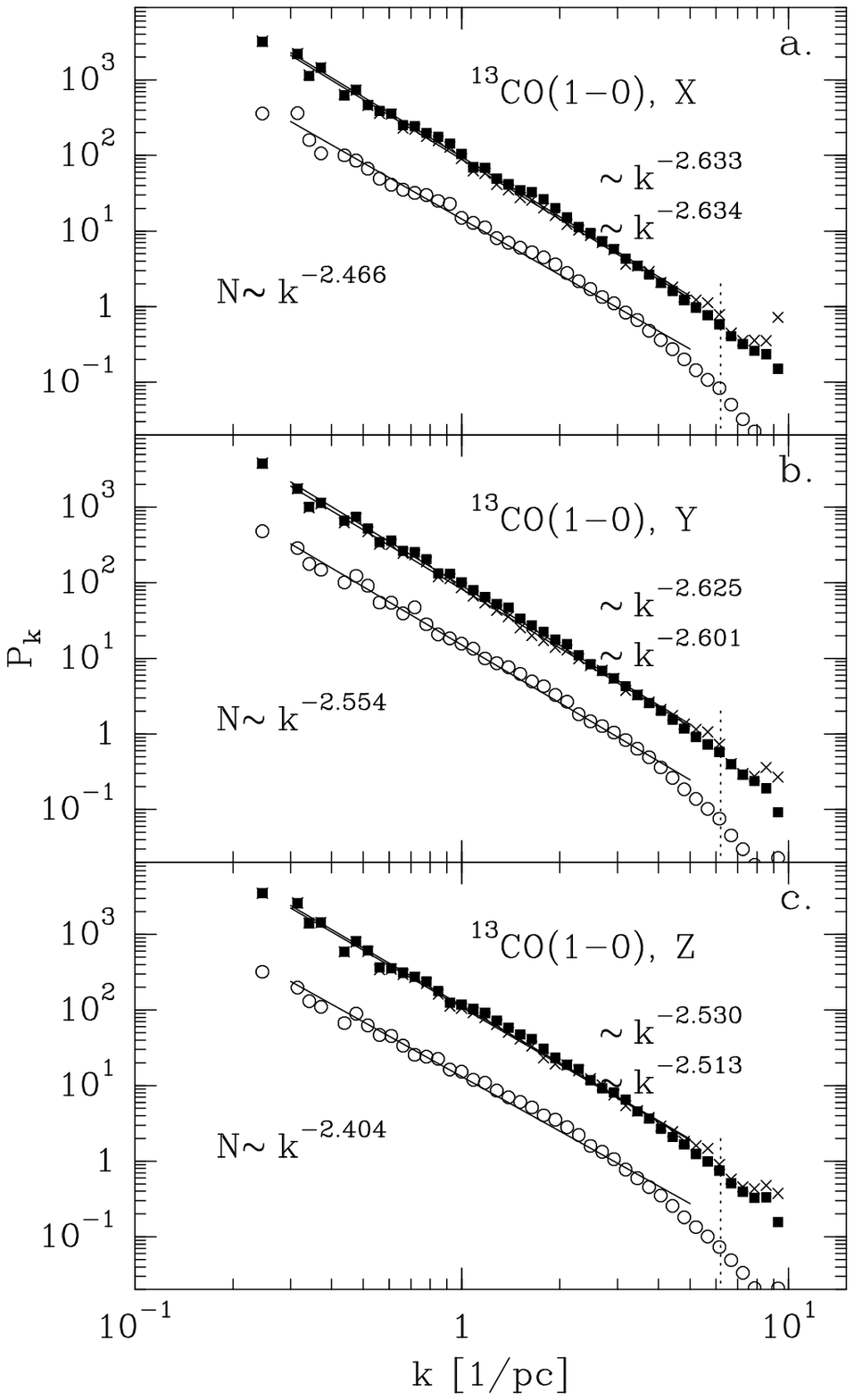}
\caption[]{
Spatial power spectra of the $^{13}$CO line intensity and of the column density in 
model $B$. Frames correspond to maps computed for three orthogonal directions. Solid
squares show the spatial power spectrum of the $^{13}$CO emission from the full 
124$^{3}$ cell model. 
Crosses show the same power spectra computed from multi-resolution models where
cells with densities above 2000\,cm$^{-3}$ are refined ($R$=0.047). Open
circles show the column density power spectra (the normalization is arbitrary). 
Solid lines indicate least squares fits in range 0.3-5\,pc$^{-1}$. The slopes 
are indicated in the figure. Slopes for $^{13}$CO are shown on the right hand 
side (the values for the multi-resolution models are below those for the full 
resolution models). The dotted vertical lines indicate the Nyquist limit for 
the full resolution model with 124$^3$ cells.
} \label{fig:power_13co_2000}
\end{figure}

Spectral line maps were computed toward three orthogonal directions using both
full resolution and multi-resolution models. The maps were used to test the
accuracy of spatial power spectra obtained with multi-resolution calculations.
Least square fits to the power spectra were computed  in the interval
$k$=0.3-5\,pc$^{-1}$. In the case of the full 124$^3$ SR model the Nyquist
limit is 6.2\,pc$^{-1}$ and, for the grid of lowest resolution (level 1 of the MR
model), one fourth of this value, 1.55\,pc$^{-1}$.
Fig.~\ref{fig:power_13co_2000} shows results for $^{13}$CO spectra from model
$B$ where the density threshold for refinement was 2000\,cm$^{-3}$. In this case, 
$R$=5\% (factor of 20 reduction in computational time and memory requirement) and 
the slope of power spectra is recovered with an accuracy of $\sim$1\%. 
This can be compared with the difference between the slopes of line
intensity and column density power spectra which is at least three times and
on the average about ten times larger.
Table~\ref{table:power} shows results for both models
($A$ and $B$) and both molecules ($^{13}$CO and $^{12}$CO) when cells above
$n=500$\,cm$^{-3}$ are refined.

\begin{table}
\caption[]{
Accuracy of power spectra from the multi-resolution models where cells above
$n$=500\,cm$^{-3}$ are refined. Columns are (1) model cloud, (2) $R$
parameter, (3) slope of spatial power spectrum of column density, (4)
molecule, (5) slope of spatial power spectrum of $^{13}$CO intensity
from the multi-resolution models. The slopes are averages of results obtained for
three orthogonal directions. The average difference between slopes computed from
the multi-resolution models and the full 124$^3$ cell models are given in parentheses 
in the last column
}
\begin{tabular}{lllll}
\hline
Cloud &  $R$   &   $a(N)$    & Molecule    &  $a(^{13}{\rm CO})$  \\
\hline
 $A$  &  0.31  &   2.831     &  $^{13}$CO  &    2.641 (-0.027)    \\
      &        &             &  $^{12}$CO  &    2.527 (+0.009)    \\
 $B$  &  0.24  &   2.475     &  $^{13}$CO  &    2.583 (+0.017)    \\
      &        &             &  $^{12}$CO  &    2.570 (-0.015)    \\
\hline
\end{tabular}
\label{table:power}
\end{table}

\section{Comparison with LTE- and non-isothermal models} \label{sect:comparison}

We have used the new ALI multi-resolution radiative transfer code to study 
differences in statistical properties between LTE and non-LTE spectra and
between isothermal and non-isothermal models. Non-LTE non-isothermal models
provide the closest descriptions to real interstellar clouds. However, 
LTE and isothermality are often assumed when MHD simulations are compared 
with observations. It is therefore important to quantify the uncertainties 
caused by these assumptions.

The cloud model used for this purpose is based on an MHD simulation performed 
on a uniform cartesian grid of 350$^3$ computational cells. In order to make the
grid divisible by four we took from this 340$^3$ cells. The simulation represents 
a cloud with a sonic Mach number $M\approx 10$ and an average density of 
$n=500$\,cm$^{-3}$ \citep[for details see][]{padoan04}. We define a three level 
hierarchical refinement, where the root grid (level 1) consists of 85$^3$ cells. 
The grid is refined above a density threshold of $n>500$\,cm$^{-3}$. After
two sub-divisions the smallest cells are the same as in the full resolution model
of $340^3$ cells. The computational time is about one quarter of that needed at 
the full $340^3$ resolution ($R\approx$0.26). Based on previous tests, the rms 
error of line intensities should be below $6\%$. However, even better accuracy is 
expected because the total optical depths are similar as in Sect.\ref{sect:tests}, 
but the resolution is higher.

\subsection{Comparison with LTE calculations} \label{sect:lte}

Line observations are often studied with an 'LTE' analysis, which includes
both assumptions of local thermodynamic equilibrium and constant excitation
along the line of sight. Conversely, under the same assumptions spectral lines
are easily predicted for any cloud model. The source function and the optical
depth follow from the LTE conditions and the radiative transfer equation can
be directly integrated along the line-of-sight. The LTE method is very often used
in the analysis of observations and also to compare observations with
MHD simulations. Although this method takes into account of some optical
depth effects, it is extremely simplified considering the wide range of
physical conditions present in any cloud.

\begin{figure}
\epsscale{0.63}
\plotone{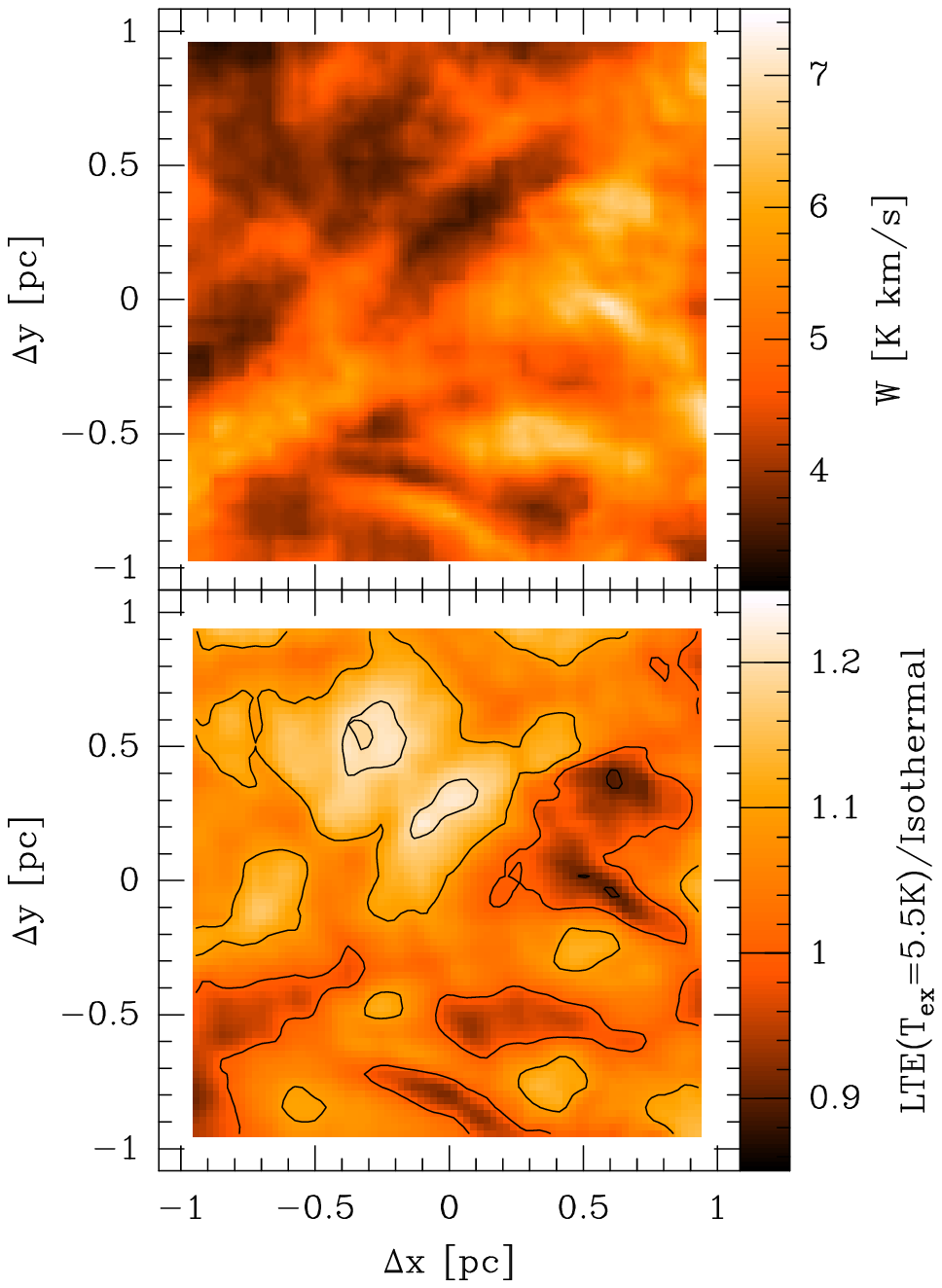}
\caption[]{
The upper frame shows a map of line intensities calculated from our non-LTE isothermal
cloud model ($T_{\rm kin}$=10\,K). The lower frame shows ratio of line intensities
between LTE-model with $T_{\rm ex}$=5.5\,K and the isothermal model.
These images are restricted to a small region of 2$\times$2\,pc, while the full 
size of the model is 10$\times$10\,pc.
} \label{fig:ratio_lte}
\end{figure}

We computed a non-LTE isothermal model with $T_{\rm kin}$=10\,K and compared it
with spectra obtained under LTE conditions and assuming constant excitation
temperature for the whole cloud. In a real cloud the excitation temperature 
varies from position to position and can take any value between the kinetic 
temperature and the temperature of the background radiation (2.7\,K). 
We considered two LTE models, one with $T_{\rm ex}=10$\,K and another with 
$T_{\rm ex}=5.5$\,K. In the following we refer to the velocity integrated line
intensity (measured in K km s$^{-1}$) simply as line intensity or line area.
We find the average line intensity of the $T_{\rm ex}$=10\,K model 
is approximately  30\% higher than in the non-LTE model, while it is approximately 
15\% lower in the $T_{\rm ex}=5.5$\,K model. After subtracting the mean intensity, 
the remaining rms intensity scatter between the LTE model with $T_{\rm ex}=5.5$\,K 
and the isothermal model is approximately 25\%. This shows that there are significant 
differences in the spatial distribution of line intensities. Intensities of the
$T_{\rm ex}=10$\,K model are compared with the LTE model in Fig.~\ref{fig:ratio_lte}. 

We computed spatial power spectra based on the $^{13}$CO(1-0) line areas.
Least squares fits to these spectra are computed in the range of wavenumbers
$k=0.3-10.0$\,pc$^{-1}$. The upper limit is well below the Nyquist limit 
(the actual resolution of MHD simulation is slightly worse than the sampling). 
The slopes of the fits are given in Table~\ref{tab:lte}. Differences are
very small, of the order of a few per cent, compared with the large line
area variations seen in Fig.~\ref{fig:ratio_lte}. 
Apparently for lower $T_{\rm ex}$ the main effect is a decrease of power at
all scales and the slope of the power spectrum is only slightly affected.

\begin{table}
\caption[]{
Slopes of spatial power spectra of $^{13}$CO line area in the $340^3$ cell non-LTE isothermal model
with $T_{\rm kin}$=10\,K and in two LTE models with $T_{\rm ex}$=10\,K and
5.5\,K. The last column gives the slopes for column density maps of the MHD model. 
Results are given for line area maps computed for three orthogonal directions ($X$, $Y$,
and $Z$). 
}
\begin{tabular}{lllll}
\hline
direction & $T_{\rm kin}$=10\,K & $T_{\rm ex}=10$\,K & $T_{\rm ex}=5.5$\,K & $N$ \\
\hline
 $X$  &  -2.621  &   -2.661   &  -2.628   &  -2.663  \\
 $Y$  &  -2.665  &   -2.691   &  -2.658   &  -2.688  \\
 $Z$  &  -2.674  &   -2.714   &  -2.642   &  -2.746  \\
\hline
\end{tabular}
\label{tab:lte}
\end{table}

Replacing non-LTE level populations with LTE values decreases the 
excitation in dense regions and increases it in the low density gas. 
This should make the spatial power spectra of the intensity maps closer to 
those of the column density. This is in fact the case of the $T_{\rm ex}$=10\,K 
model, where the slopes of the power spectra of line areas are almost identical 
to the slopes of the column density maps, within the numerical accuracy. 
The accuracy of the computed slopes should be better than $\sim$0.01 
(see Table~\ref{table:power}) because the same multi-resolution cell
discretization is used for the three models. For excitation temperature $T_{\rm
ex}$=5.5\,K the slopes are closer to those of the isothermal model and, apart
from the $X$-direction, slightly lower. The overall differences between the
three clouds are small and mostly reflect average excitation temperatures. For
such cold clouds small temperature differences correspond to large changes in
optical depth. At $T_{\rm ex}$=5.5K the optical depth of the $J=1-0$ line is
already more than twice the value at $T_{\rm ex}$=10\,K. Larger differences between
line maps and column density maps are therefore expected for the $T_{\rm ex}$=5.5K
model than for the $T_{\rm ex}$=10\,K one.

This comparison of LTE and non-LTE spectra is different from the issue of the
accuracy of LTE column density estimates \citep[see e.g.][]{padoan00}.
\citet{padoan04} analysed simulated $^{13}$CO spectra and concluded that the
power spectrum of column densities derived from observations under the LTE
assumption was generally steeper than the power spectrum of the underlying
column densities. One reason is that as $^{13}$CO lines become saturated in the
densest parts of the cloud the column densities are underestimated there
and also the power at small scales is reduced. The effects of such
self-absorption are directly visible in line profiles only for optically
thicker species. 

Another reason may be linked with excitation temperature gradients. In LTE
calculations $T_{\rm ex}$ is usually based on the brightest regions where
$T_{\rm ex}$ is above the average value. This means that in regions of lower
density (and lower $T_{\rm ex}$) column densities are underestimated. If this
introduces additional gradients between centre and edges of the column density
map this will add power at large scales and make the power spectrum steeper.
On the other hand, the same effect can work also at small scales. In
particular, if the excitation temperature in a core is higher than assumed the
computed column density will also be too high and additional power is
introduced at the smallest scales.

\subsection{Comparison with non-LTE non-isothermal models}

\begin{figure}
\epsscale{0.62}
\plotone{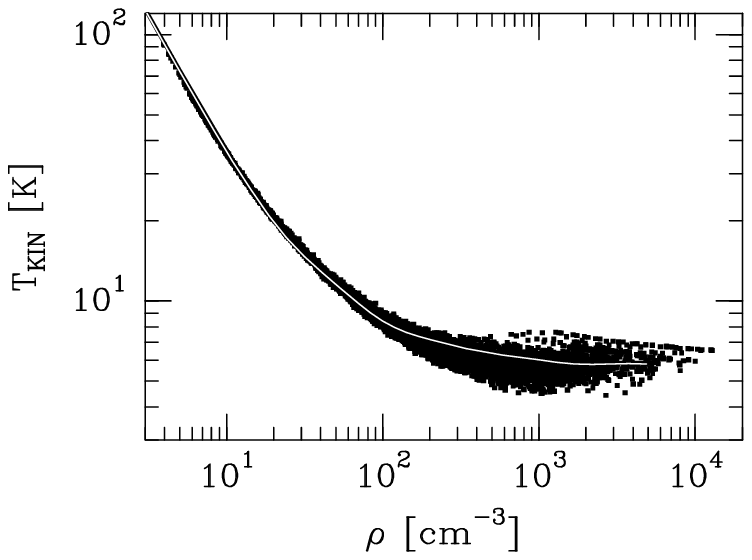}
\caption[]{
Kinetic temperature as the function of local density \citep[for details
see][]{padoan04}. The solid line represents a fit that was used to set the kinetic
temperature in our non-isothermal model. 
} \label{fig:tkin}
\end{figure}

\citet{padoan04} computed kinetic temperatures for similar model
clouds by balancing cosmic ray heating with cooling by line emission from 
the most important species. This resulted in a relation between the gas kinetic
temperature (see Fig.~\ref{fig:tkin}). We use this relation to compute 
non-LTE non-isothermal spectra for our 340$^3$ cell model. These spectra should
provide a better description of observational data than the isothermal ones.

\begin{figure}
\plotone{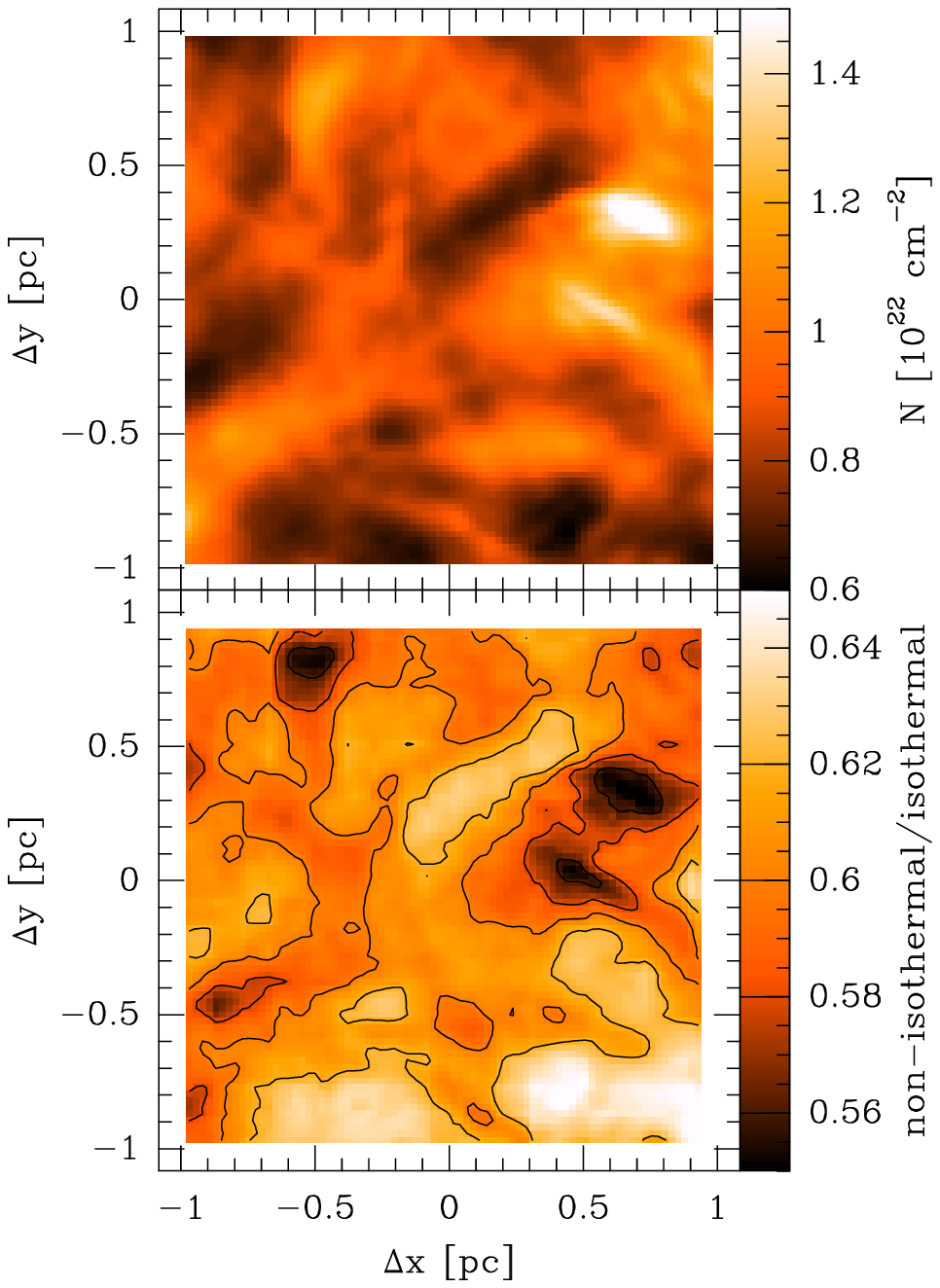}
\caption[]{
The upper frame shows the column density around the centre of the model cloud. The
lower frame shows the ratio between $^{13}$CO(1-0) line areas in non-isothermal and 
isothermal models. The images are limited to a region of 2$\times$2\,pc. The full 
size of the model is 10$\times$10\,pc.
} \label{fig:ratio_t}
\end{figure}

Because the kinetic temperature decreases as a function of density one would
expect emission from dense cores to be decreased, as in the LTE models. This 
is to some extent visible in Fig.~\ref{fig:ratio_t}, where cores with the brightest 
$^{13}$CO emission are weaker in the non-isothermal case than in the isothermal one. 
However, differences from the LTE model (Fig.~\ref{fig:ratio_lte}) are also significant,
due to the large range of kinetic temperatures in the non-isothermal cloud.

\begin{figure}
\plotone{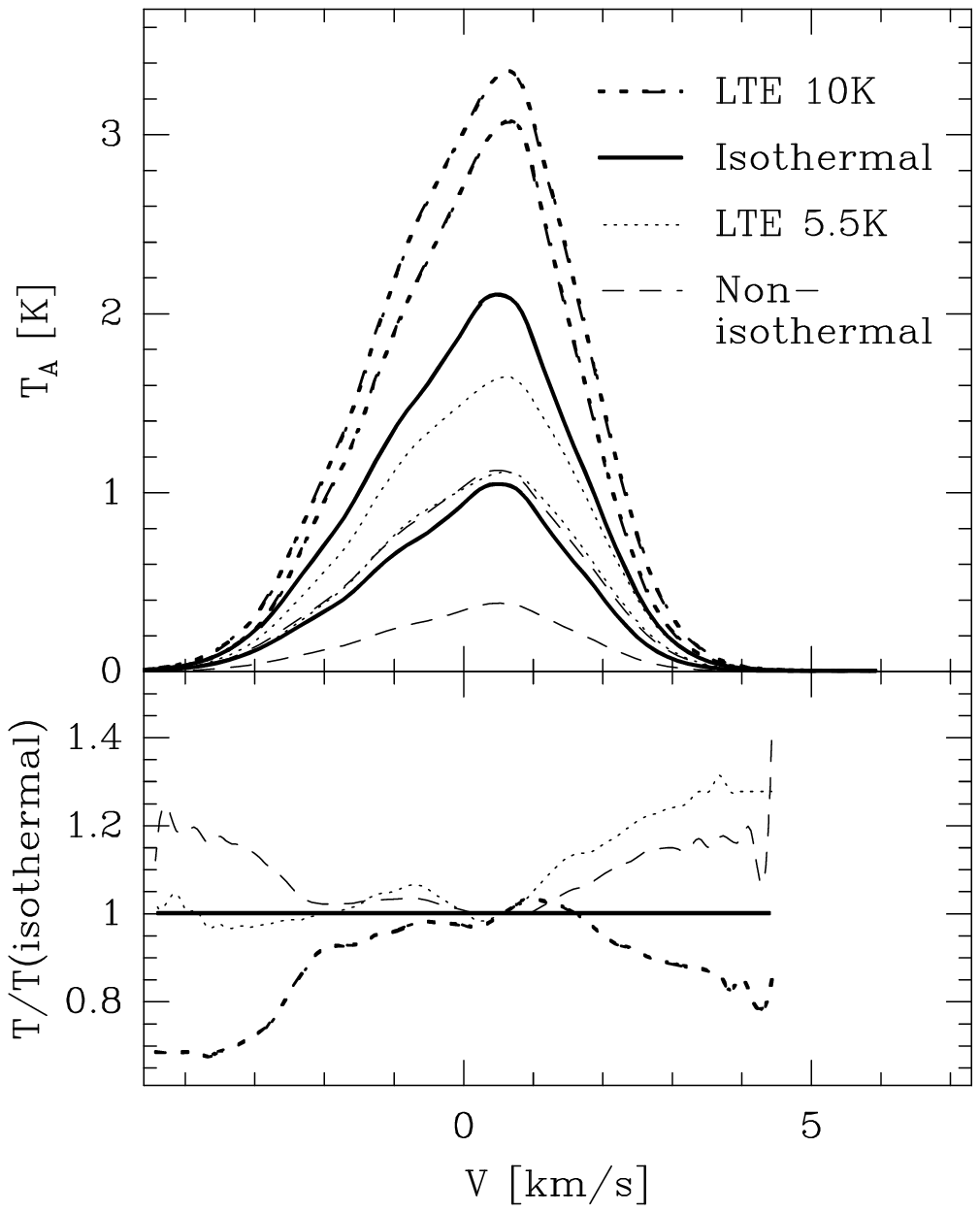}
\caption[]{
The upper panel shows the average $^{13}$CO(1-0) and $^{13}$CO(2-1) 
spectra from the isothermal model, the
non-isothermal model and the LTE model of Sect.~\ref{sect:lte}.
The $J=2-1$ line is stronger than $J=1-0$ line only in the case of LTE model with
$T_{\rm kin}$=10\,K. The spectra are averaged over the whole spectral line map
computed for the $X$ direction. 
The lower panel shows $^{13}$CO(1-0) line profiles normalized to the same peak
value and divided by the average profile of the isothermal model.
} \label{fig:ave_spectra}
\end{figure}


Fig.~\ref{fig:ave_spectra} shows average $^{13}$CO(1-0) and $^{13}$CO(2-1)
spectra. The profiles of the averaged spectra are smooth but significant
differences are seen in the line ratios between the LTE and the non-LTE
models.  Differences are also visible in $^{13}$CO(1-0) profile ratios plotted
in the lower panel of Fig.~\ref{fig:ave_spectra}. Compared with the isothermal
cloud the LTE model with $T_{\rm ex}$=10\,K has weaker line wings while for
the LTE model with $T_{\rm ex}$=5.5\,K the opposite is true. The optical depth
of the cold LTE model is much higher than in the warmer LTE model, so the
profile reflects more the velocity structure on the observer side of the
cloud. The non-isothermal model shows stronger line wings that result from the
fact that in this model kinetic temperatures are higher in the more rapidly
moving low density gas. This should also be the case for observed spectra of
molecular clouds, as similar kinetic temperature variations should also be
present in real clouds.

\begin{figure}
\plotone{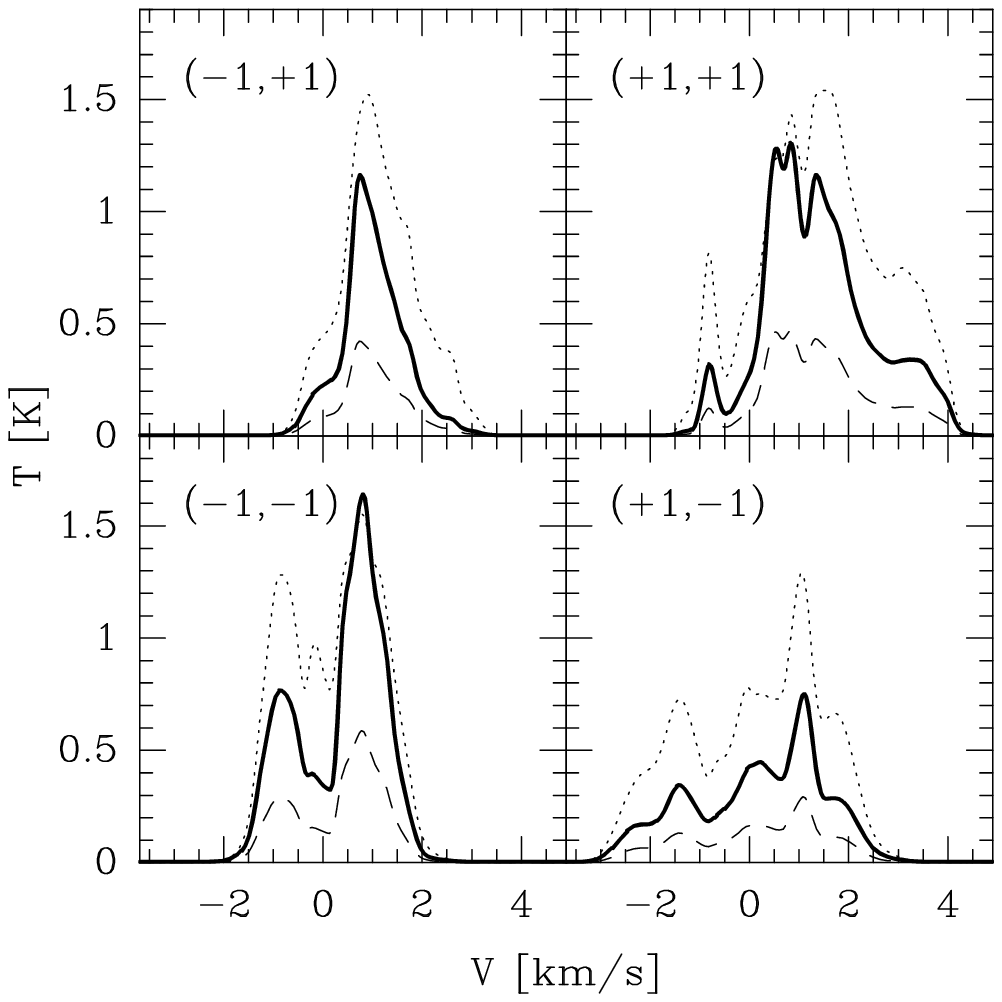}
\caption[]{
A few $^{13}$CO(2-1) line profiles computed for non-LTE isothermal and non-isothermal
models and for the LTE model with $T_{\rm ex}$=5.5\,K. Spectra are calculated toward
the $X$ direction and averaged over an area of 0.04\,pc$^2$. Positions are given in 
the figure as offsets in parsecs from the map centre. These positions correspond to the 
corners of the map shown in Fig.~\ref{fig:ratio_t}.
} \label{fig:profiles}
\end{figure}

Differences between models are even stronger in the profiles of individual spectra,
as shown in Fig.~\ref{fig:profiles}. In some velocity intervals, spectra may be almost
identical, while at other velocities differences between the LTE and the non-LTE 
non-isothermal models can exceed a factor of four.
Despite the differences in the peak temperature of non-isothermal and isothermal
models (over a factor of two) the shape of the spectral line profiles are similar
in these two models. Spectra from the LTE model show both different peak temperatures
and spectral shapes relative to the non-LTE models.

\begin{figure}
\plotone{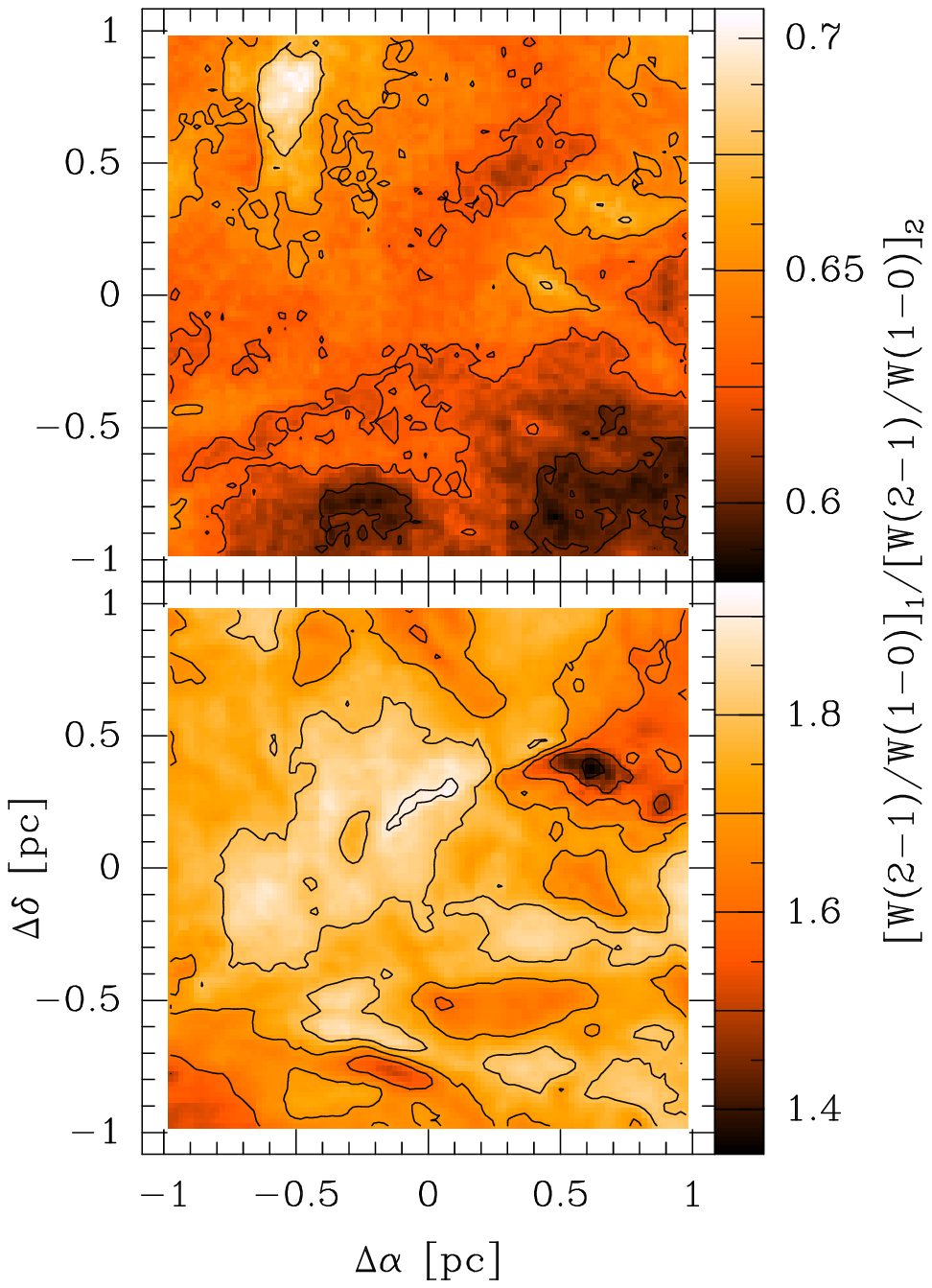}
\caption[]{
Line ratios $^{13}$CO(2-1)/$^{13}$CO(1-0) in relation to the isothermal model.
The upper frame shows the ratio of integrated line areas in the non-isothermal model
divided by the ratio in the isothermal model. The lower frame shows line
ratios from the LTE model with $T_{\rm ex}$=5\,K divided by line ratios
from the isothermal model. Only the central region of 2$\times$2\,pc is shown.
} \label{fig:line_ratio_ratio}
\end{figure}

Fig.~\ref{fig:line_ratio_ratio} shows ratio of line areas,
$W$[$^{13}$CO(2-1)]/$W$[$^{13}$CO(1-0)], for non-LTE-non-isothermal and LTE models
divided by the same line area ratios in the isothermal model. In both cases, variations 
up to 40\% are found.

Differences in the spatial power spectra calculated based on the
$^{13}$CO(1-0) line areas are very small, with the slopes of the
non-isothermal model steeper by only $\sim$1\%. The shape of the $T_{\rm kin}$
vs $n$ relation is such that the model cloud is almost isothermal at the
highest densities. Some of the differences between the isothermal and
non-isothermal models may well be caused by a difference in the average
kinetic temperature, which affects the optical depths. In the case of the
LTE models we found a flattening of the spatial power spectra with decreasing 
excitation temperature (increasing optical depth). The spatial power
spectra are flatter for the non-isothermal model which is consistent with the
average kinetic temperature being below 10\,K.

The correlation between kinetic temperature and local gas density plotted
in Fig.~\ref{fig:tkin} contains a significant amount of scatter. There is some
noise arising from the used Monte Carlo calculations. However, most of the 
scatter is real, showing the kinetic temperature does not depend
only on the local gas density but also on the shielding
provided by neighbouring regions. In particular, the kinetic temperature
distributions as calculated in \citet{padoan04} have large
scale structure with temperature decreasing toward centre of the cloud. When
we use the fit shown in Fig.~\ref{fig:tkin} we ignore this variation and
consequently we subtract power from the large scales. 

Such large scale gradients affect also the reverse process of estimating
column densities from observations using the LTE analysis. When a constant
excitation temperature is determined from the dense central regions the
excitation is progressively over-estimated as one moves toward the cloud
edges. At the same time the ratio between estimated column density and the
true column density falls below one. In other words, column density gradients
between centre and cloud edge are artificially increased, more power is
introduced at large scales and one obtains a spatial power spectrum that is
too steep (i.e. steeper than the corresponding spectrum of true column
densities).  This may be the reason why \citet{padoan04} found a large difference
between the slopes of LTE column density estimates and true column density.
They estimated this difference to be 0.13$\pm$0.06. In our case differences
between different cases are about 0.04 or below but by construction our models
do not have similar $T_{\rm kin}$ gradients. We should again point out that
our comparison between LTE and non-LTE models is different from the problem of
estimating LTE column densities from observations. However, if our
interpretation is correct, the effects that such large scale temperature
gradients (in $T_{\rm ex}$ amplified by $T_{\rm kin}$ gradients) have on
spatial power spectra slopes can be as significant as differences between e.g.
LTE and non-LTE calculations.

\section{Discussion}   \label{sect:discussion}

Comparison of LTE and non-LTE models showed significant differences in line
profiles but differences in spatial power spectra were relatively small. The
slopes of the power spectra depend, however, on the assumed excitation
temperature since that determines the line optical depths. For example, in
Table~\ref{tab:lte} at high temperatures (low optical depths) the LTE slope
approaches that of the column density distribution and at low temperatures
(high optical depths) the spectrum becomes more shallow than in non-LTE
calculations. If one is interested {\em only} in the slope of the power
spectrum and an accuracy of a few per cent is sufficient then LTE calculations
can be used to predict line intensities. When results are compared with
observations it is also better to use LTE predictions rather than column
densities read directly from the cloud model. In our models the differences
between power spectra of column density and non-LTE line intensities were less
than 3\%. Therefore, if one considers only power spectrum slopes then non-LTE
calculations are necessary only in quite detailed studies. However, they are
still required if absolute line intensities, line ratios or line profiles are
being studied.

The spatial power spectra were fitted with power laws $P\sim k^{-a}$. The slopes
$a$ were found to depend on two factors, the average optical depth and the
general cloud structure. In most cases (models $A$ and the large 340$^3$ cell
models) the power spectrum is flatter for line emission than for the
underlying column density distribution. Furthermore, $a$ decreases with
increasing optical depth, as seen by comparing the results for
$^{13}$CO and $^{12}$CO. However, if the slope of the column density
spectrum is shallow ($a$ below 2.5) the spectrum of line emission can become
the steeper of the two and the effect of optical depth on $a$ is reduced
or may even be reversed. For model $B$ all $^{13}$CO maps and two of the three
$^{12}$CO maps have power spectra that are steeper than the spectra of the
corresponding column density maps. For the direction $z$ the slope $a$ is also
steeper for the $^{12}$CO map than for the $^{13}$CO map. The trend is clear:
as the slope of the column density becomes more shallow the difference between
$^{12}$CO and $^{13}$CO maps decreases (see Table~\ref{table:slopes}) and only
in the case of the smallest slope ($a\sim$2.4) the spatial power spectrum of
the optically thicker species is the steeper one.

\begin{table}
\caption[]{Comparison of spatial power spectra computed for $^{12}$CO and
$^{13}$CO maps. The non-LTE spectra were computed for isothermal models with
$T_{\rm kin}$=10\,K. The columns are: the model cloud, the viewing direction,
slope of the column density power spectrum ($a_{\rm N}$), slope of the spatial
power spectrum for $^{13}$CO emission and the difference in the slopes between
$^{12}$CO and $^{13}$CO power spectra. Maps are listed in increasing order of
the slope $a_{\rm N}$.  }
\begin{tabular}{lllll}
\hline
model & direction & $a_{\rm N}$   & $a(^{13}{\rm CO})$  &  $a(^{12}{\rm CO})-a(^{13}{\rm CO})$ \\
\hline
$B$  &  $z$   &   2.40  & 2.51  & +0.11  \\
$B$  &  $x$   &   2.47  & 2.63  & -0.06  \\
$B$  &  $y$   &   2.55  & 2.60  & -0.08  \\
$A$  &  $z$   &   2.71  & 2.61  & -0.05  \\
$A$  &  $x$   &   2.85  & 2.65  & -0.12  \\
$A$  &  $y$   &   2.93  & 2.66  & -0.16  \\
\end{tabular}
\label{table:slopes}
\end{table}

\begin{figure}
\plotone{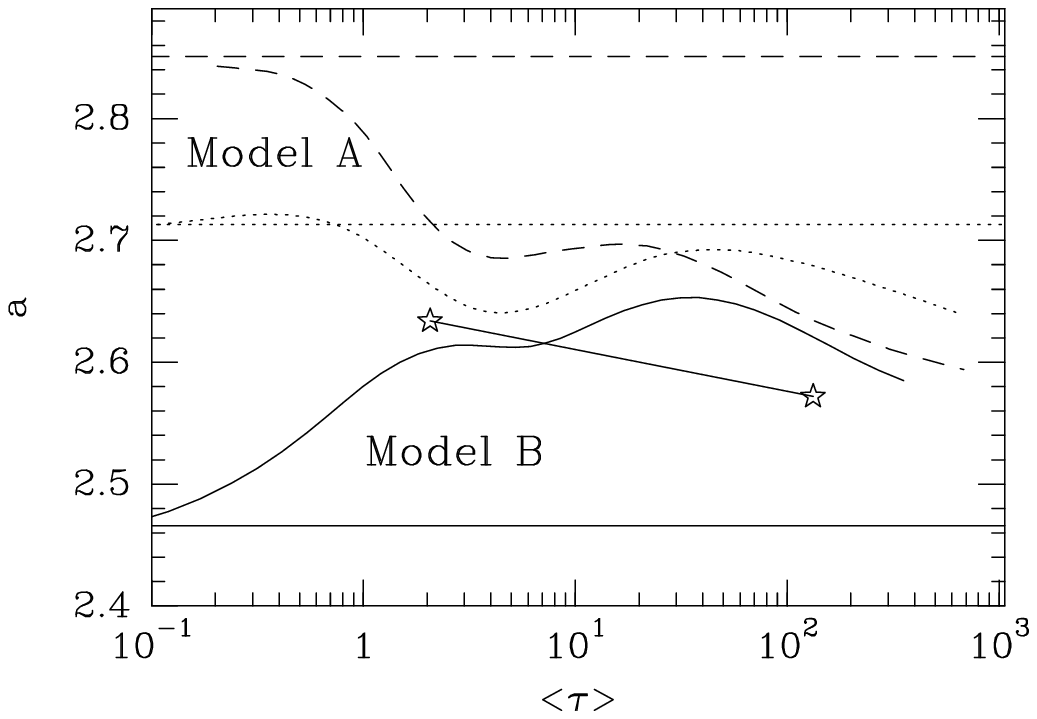}
\caption[]{
Slopes of the power spectra as a function of average optical depths. The
curves show the behaviour for direction $x$ in model $B$ (solid line) and for
model $A$ in directions $x$ (dashed line) and $z$ (dotted line) according to
LTE calculations with $T_{\rm ex}$=10\,K. The horizontal lines indicate the
corresponding slopes for the column density maps. The stars show values for two
non-LTE clouds ($T_{\rm kin}=10$\,K) based on model $B$.
} \label{fig:taus}
\end{figure}

In the optically thin limit the slopes of line emission and column density
power spectra are identical. As the optical depth increases the densest cores
are the first ones to become optically thick and emission from those regions 
saturates. Assuming the cores are very compact (as in model $B$) power is
removed from the smallest scales and the spatial power spectrum of line
emission becomes steeper. Clearly this explanation does not apply e.g. to
model $A$ since there the spectrum becomes more shallow instead. The
actual optical depth dependence can be quite complicated and this can be seen
even from LTE models not including effects that result from spatial variations
of the excitation temperature. Fig.~\ref{fig:taus} shows the change in the 
slope of the spatial power spectrum as a function of the average optical 
depth in models $A$ and $B$. For these LTE models a constant excitation temperature
of 10\,K was assumed and the molecular abundance was scaled to produce clouds
with different optical depths. 

In non-LTE calculations for model $A$ the spatial power spectrum of line 
emission is less steep than the spectrum of the corresponding column 
density map. The opposite is true for model $B$. This is also
illustrated in Fig.~\ref{fig:taus},
which shows results from LTE models with a wider range of optical depths.
Differences between column density and line maps increase at least up to
$<\tau>\sim 100$ when most of the sightlines are already optically thick. 
However, the change is non-linear and depends on the actual density distribution
and on the order in which different regions become optically thick. Furthermore,
the $a$ vs. $\tau$ curves can be very different even when the same model cloud
is viewed from different directions.

The slopes from non-LTE calculations with model $B$ are marked in
Fig.~\ref{fig:taus}. For $^{13}$CO map ($<\tau>=2.1$) the slope is steeper
than when LTE conditions were assumed.  In our non-LTE model the excitation
temperature is lower than in the LTE-model and optical depth of the $J=1-0$
transition tends to be higher. As a result, in the figure the mass of the
non-LTE cloud is smaller than the mass of a LTE cloud with the same average
optical depth.  The non-LTE point corresponding to $^{12}$CO map
($<\tau>=133$) is again below the curve of LTE models. At least in this range,
the optical depth dependence of non-LTE models is different from that of LTE models.

The tests of Sect.~\ref{sect:tests} showed that multi-resolution models can
provide significant savings in computational time and memory. An accuracy of
approximately 10\% was obtained in $^{13}$CO calculations with $R\sim$0.1.
This corresponds to a factor of ten reduction in computational time and memory
requirements (see also Appendix~\ref{sect:implementation}). 

In this paper we have computed spectral line maps over entire model clouds.
Multi-resolution models provide, however, a general framework with which we
can easily direct our attention to selected parts of a larger model cloud. For
example, if we are interested only in one core we can use a discretization
that becomes more coarse as distance between our line-of-sight and the core
increases. In other words, long distance interactions between the core and the
rest of the cloud are all included but are treated in a more approximate
way. This can lead to very low $R$ values ($\sim$1\% or below)
making such radiative transfer calculations feasible for parts of future large
MHD simulations (such as simulations of core collapse in large
turbulent clouds computed with AMR methods).

\section{Conclusions}  \label{sect:conclusions}

We have presented a radiative transfer code for line transfer calculations
on hierarchical grids. In interstellar clouds, dense gas responsible for 
the observed molecular lines has a small volume filling factor. As a result,
fine discretization is needed only in a limited sub volume and hierarchical 
grids can significantly reduce both
time and memory needed for radiative transfer simulations. The code was
tested using cloud models that were based on MHD simulations. In particular we
looked at the parameter $R$ which is the ratio between number of cells in the
hierarchical grid and in the corresponding single resolution grid (equal size of
the smallest cells). According to these tests:
\begin{itemize}
\item for lines with optical depths $\tau\sim$1 a 10\% accuracy of computed line
intensities is reached with $R \sim$0.1
\item for optically thick lines ($\tau \ga$100) an accuracy of 10\% is reached
with $R \sim$0.15
\item if higher accuracy is needed the required number of cells increases
rapidly: in the case of high optical depths a 5\% accuracy of line intensities
requires already $R>0.5$ and the advantage of hierarchical grids becomes small 
\item the spatial power spectra recovered from hierarchical models were found
to be accurate: with $R=0.05$ errors in the slopes of the power spectra were 
only $\sim$1\%
\end{itemize}

Using the new radiative transfer program we computed $^{13}$CO spectra for one
model cloud (340$^3$ cells) first for LTE conditions and then with full
radiative transfer calculations assuming either isothermal or non-isothermal
conditions. The comparison of the results showed that
\begin{itemize}
\item the LTE models differ clearly from the non-LTE models - not only
in line intensities but especially in the shape of the line profiles
\item the differences in the slopes of spatial power spectra, $a$, are small
($\la$2\%) and the main effect is a slight steepening of the power spectra with
increasing excitation temperatures
\item there is some indication that large scale gradients in kinetic
temperature between cloud centre and surface (not included in our models here)
may produce a steepening of the power spectrum that is larger
than the differences between the models considered in this paper
\item the difference between spatial power spectra of column density and line
intensities was found to depend on both optical depth and cloud structure
\item in models with relatively little small scale structure ($a\ga2.5$)
the slope of the spatial power spectrum is more shallow for line emission
than for column density and becomes even flatter with increasing optical depth
\item in a more `clumpy' cloud ($a\la2.4$) the situation may be reversed:
line emission has a steeper power spectrum and slope increases with optical
depth

\end{itemize}

\begin{acknowledgements}
M.J. acknowledges the support of the Academy of Finland Grants no. 
175068, 174854, 1201269, and 1206049. 
\end{acknowledgements}

\appendix


\section{Implementation of multiresolution calculations}
\label{sect:implementation}

The implementation presented in this paper is optimised for low memory
requirement. This is in part achieved by simulating each transition 
separately, and by writing intensities at cell centres to external files
before proceeding to simulation of the next transition. The separate handling
of transitions introduces an overhead that can be removed at the cost of an
increased memory consumption. In our scheme, the intensities of the incoming
radiation are kept in memory for one layer of level 1 cells and their
sub-cells. These intensity values correspond to a significant fraction of
total memory requirements. If all transitions were simulated simultaneously,
memory requirement would increase by a factor almost equal to the number of
transitions. On the other hand, because intensities are needed only for one
layer of cells, the size of these arrays increase only proportionally to $L^2$
where $L$ is the number of cell layers in the full-resolution model (i.e. full
resolution model would contain $L^3$ cells).  The use of external files for
centre intensities (one value per transition per cell) may be more significant
for the run times.  These values can be kept in main memory only for small
models since their number increases proportionally to $L^3$. 

For illustration we can look at the memory requirements of the computations
presented in Sect.~\ref{sect:comparison} where $L$ was 340. One cell layer
consists of $(L/4)^2$ level 1 cells, each containing down to level 2 a maximum
of 64 sub-cells. If each transition is treated separately and intensities are
simulated using $\sim$100 frequency points the maximum storage requirement
equals $\sim (340/4)^2\times64\times100$ floating point numbers i.e. some
180MB. The space needed to store average intensities in cells is $R\times L^3
\sim 0.25 \times 340^3$ numbers which equals about 40MB per transition. Additional
space is needed for storing basic cloud data (cell velocities, densities,
doppler widths etc.) and data needed for simulation of one transition (optical
depths, source functions). These amount to $\sim$300MB but are mainly kept
in external files.


For a cloud consisting of a regular grid of level 1 cells there is no overhead from
the machinery that exists for handling multiresolution grids. For hierarchical
grids the computations do, however, become somewhat more complicated. We take
as an example model $B$ where cells above density limit $n=500$\,cm$^{-3}$
were divided and the value of $R$ was 0.24. Compared with full grid ($124^3$
level 1 cells) the run time dropped to 0.27 times the original. This is quite
close to that actual value of $R$ and shows that it is a good indicator
of the reduction in run times.

We have computed multi-resolution models with a maximum of three levels in the
cell hierarchy. Deeper hierarchies are needed in the case of large MHD
simulations on uniform cartesian grids with $L\sim 1000$ or for MHD
simulations with AMR methods, where grids with many levels of refinement can
be generated.  In our current scheme each individual cell can be sub-divided
independently. This is optimal as far as the $R$ values is concerned but may
become inefficient for deeper hierarchies. Therefore, an alternative gridding
scheme is being developed which will use continuous rectangular regions
with fixed cell sizes. There the hierarchy will consist not of subdivided
cells but of sub-volumes with smaller cell sizes.


\end{document}